\documentclass[twocolumn,pre,showpacs,showkeys]{revtex4-1}
\usepackage{graphicx}
\usepackage{soul}
\usepackage{xcolor}
\usepackage{amsmath}
\usepackage{pgf}
\usepackage{float}
\usepackage[greek,english]{babel}
\usepackage{subcaption}
\usepackage{caption}
\usepackage{amssymb}

\begin{document}

\title{Nonlinear dynamics and chaos in dissipative optical trimers with complex couplings}

\author{Johanne Hizanidis}
\affiliation{Institute of Electronic Structure and Laser,
Foundation for Research and Technology-Hellas,
70013 Heraklion, Greece}

\author{Konstantinos G. Makris} 
  % \email{makris@physics.uoc.gr}
\affiliation{ITCP-Department of Physics, University of Crete, Heraklion, Greece}
\affiliation{Institute of Electronic Structure and Laser,
Foundation for Research and Technology-Hellas,
70013 Heraklion, Greece}
\date{\today}

\begin{abstract} 
In the context of non-Hermitian photonics, we consider a nonlinear optical trimer
with three lossy waveguides with complex couplings. This non-Hermitian trimer 
exhibits stable stationary and oscillatory regimes in a wide range of values of the coupling-loss parameters. Moreover, chaotic dynamics through period-doubling is confirmed via Lyapunov exponent measurements and the underlying bifurcation structure is found by semi-analytical continuation of solutions. The interplay of chaos, due to Kerr nonlinearity, and non-Hermiticity, due to the combined dissipation and complex coupling, provides a different perspective in the area of nonlinear and non-Hermitian optics.

\end{abstract}

\maketitle
%%%%%%%%%%%%%%%%%%%%%%%%%%%%%%%%%%%%%%%%%%%%%%%%%%%%%%%%%%%%
%%%%%%%%%%%%%%%%%%%%%%%%%%%%%%%%%%%%%%%%%%%%%%%%%%%%%%%%%%%%
\section{Introduction}

The collective behavior of nonlinear arrays of coupled optical waveguides has been the focus
of intense research for many decades now, as they are important for various applications in integrated 
photonics~\cite{CHR88,CHR03}. Based on the seminal work of Christodoulides \cite{CHR88}, the concept of photonic lattice was introduced to optics, and offered a paradigmatic physical system, that on the one hand played a crucial role on integrated nonlinear optics and on the other hand provided a platform for many condensed matter physics-inspired emulations \cite{CHR03}. 

The basic photonic component of such one-dimensional arrays, namely the nonlinear coherent coupler, is an analytically solvable model in terms of elliptic functions, where self-trapping of an optical beam can take place after a critical power threshold ~\cite{JEN82}. The integrable character of the nonlinear directional coupler is lost, for higher number of channels. In particular, its natural spatial extension, namely the nonlinear optical trimer (three coupled waveguides), is non-integrable and chaotic \cite{FIN90}.

Moving to lattices with even more elements, chaotic behavior has also been observed as a result of the competition between nonlinearity and linear coupling between adjacent sites in a similar system, namely the discrete self-trapping equation or the discrete nonlinear Schrodinger equation. This nonintegrable equation is highly relevant in many different areas of physics, but especially in condensed matter physics ~\cite{EIL85,DEL91} and nonlinear optics ~\cite{CHR88,CHR03}. Moreover it exhibits chaos for both periodic boundary conditions~\cite{EIL85} and nearest neighbor interactions~\cite{DEL91}, where the system's reduced symmetry, as well as, the Lyapunov exponents were systematically addressed. Note that, none of these systems exhibit non-dissipative chaos, since they are conservative (the total power is conserved) and their dynamical behavior depends on the choice of initial conditions.

In addition to nonlinearity, it has been shown that the presence of complex-valued elements in such type of systems, has been related to various effective ways to model complex laser systems and optical fiber amplifiers~\cite{CHE92}, usually in the context of discrete Ginzburg-Landau equations and dissipative spatiotemporal solitons~\cite{EFR03,EFR07}. Even though, these studies were dealing with open systems, their focus was more on nonlinear dynamics and modal engineering, rather the rich underlying nature of non-Hermiticity and symmetries.

Based on these grounds, it was not until recently that the ideas related to non-Hermiticity, where introduced to optics ~\cite{PT1,PT2,PT3,PT4,PT5,PT6} based on initial inspirations from mathematical physics concepts like the $\mathcal{PT}$-symmetry~\cite{Bender1,Bender2}. The experimental realizations of such systems requires a unique spatial combination of gain and loss materials, that are exactly balanced, in a single optical structure. This is physically possible in optics, and has triggered many theoretical and experimental works that led to the formation of the new area of non-Hermitian photonics ~\cite{PT7}. Many intricate and counter intuitive phenomena including exceptional points, unidirectioanl invisibility, ultrasensitive sensors and gyroscopes~\cite{PT8,PT9,PT10,PT11,PT12,PT13,PT14}, as well as, constant-intensity waves~\cite{CI1,CI2}, and extreme transient growth and pseudospectra of non-normal lattices~\cite{Tr1,Tr2,Tr3}.

In this framework, even though most studies were devoted to linear systems, the inclusion of Kerr nonlinearity, relevant to self-trapping, and saturable nonlinearity, relevant to lasing, has attracted recently a lot of attention. More specifically, from, $\mathcal{PT}$-solitons~\cite{PT3,NL1,NL2,ALE14,KOM16}, $\mathcal{PT}$-nanolasers~\cite{NL3,NL4,NL5,NL6,NL7,NL8}, and nonlinear control of topological phase transitions~\cite{NL9}, to asymmetric couplings of skin lasers~\cite{NL10}, and skin solitons~\cite{NL11}, optical nonlinearities have played a pivotal role to nonlinear non-Hermitian physics. 

In this paper, we study the nonlinear optical trimer with equal dissipation per channel and complex coupling coefficients between nearest neighbors (actively-coupled-AC for convenience, from now on). This model is non-integrable and thus exhibits chaos. The interplay between nonlinearity and non-Hermiticity, as well as, the transition to chaos is the focus of our work. Our theoretical results are presented as follows: In the next section, Sec.~\ref{sec:model}, we describe the model for the AC trimer with a schematic figure of the proposed physical system. In addition, we solve the system analytically in terms of its stationary solutions and compare with the numerically obtained results. Note that part of the calculations are included in the Appendix for the sake of easier readability. In Sec.~\ref{sec:chaos} we focus on the periodic solutions and the chaotic behavior of the system. Our numerical analysis is complemented by semi-analytical calculations of the underlying bifurcation structure. Finally in the conclusions, we summarize the main findings and discuss open problems for future studies.

%%%%%%%%%%%%%%%%%%%%%%%%%%%%%%%%%%%%%%%%%%%%%%%%%%%%%%%%%%%%
%%%%%%%%%%%%%%%%%%%%%%%%%%%%%%%%%%%%%%%%%%%%%%%%%%%%%%%%%%%%
\section{The model and stationary solutions}
\label{sec:model}
\begin{figure}
\includegraphics[width=0.35\textwidth]{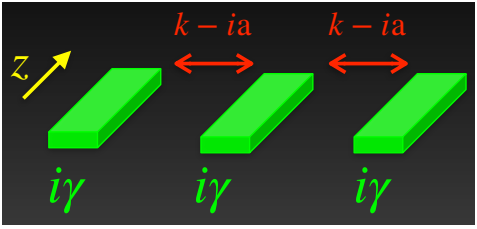}
\caption{Schematic figure of the AC trimer: three parallel lossy waveguides coupled 
via complex coupling coefficients.}
\label{fig:ac_trimer}
\end{figure}

Let us consider three parallel identical optical waveguides with loss rate $\gamma>0$ and complex coupling with 
real part $k>0$ and imaginary part $\mathrm{a}>0$, shown in the schematic Fig.~\ref{fig:ac_trimer}.
By using the coupled-mode theory approach, the optical-field dynamics in the
three coupled waveguides is described by the following set of nonlinear
equations:
\begin{eqnarray}
   \frac{d{\psi_1}}{dz}&=&-\gamma\psi_1+(ik+\mathrm{a})\psi_2 +i |\psi_1|^2\psi_1  \label{eq:ac_trimer_1}\\ 
   \frac{d{\psi_2}}{dz}&=&-\gamma\psi_2+(ik+\mathrm{a})(\psi_1+\psi_3)+i |\psi_2|^2\psi_2  \label{eq:ac_trimer_2}\\ 
   \frac{d{\psi_3}}{dz}&=&-\gamma\psi_3+(ik+\mathrm{a})\psi_2+i |\psi_3|^2\psi_3 \label{eq:ac_trimer_3},
   \end{eqnarray}
where $\psi_j$ ($j=1,2,3$) are the complex peak amplitudes of the envelope of the corresponding
electric fields in the three waveguides, respectively, and $z$ the propagation distance. Before we continue, we note here that due to the presence of the lossy terms and the complex coupling coefficients on the above equations, the corresponding linear problem is non-Hermitian. As a result, the measurable physical quantity of interest is the sum of individual intensities, namely the optical power of light in the trimer $P=|\psi_1|^2+|\psi_3|^2+|\psi_3|^2$. This optical power is a conserved quantity for both the linear and the nonlinear hermitian problem. Here it is in principle, a function of the propagation distance $z$, and plays a crucial role in our study.

\begin{figure}
\includegraphics[width=\linewidth]{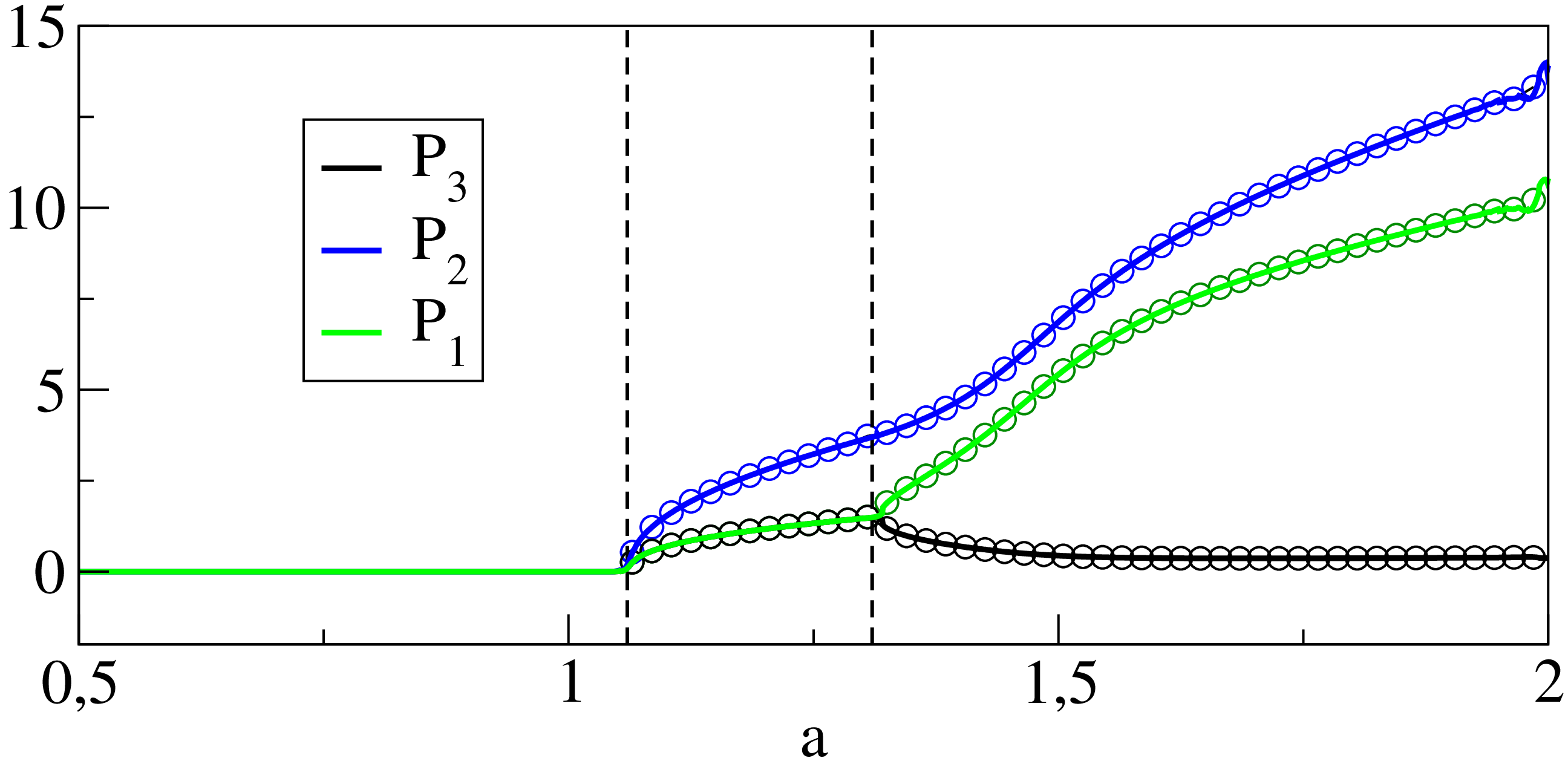}
\caption{Stationary solutions of the system of Eqs.~\ref{eq:ac_trimer_1}-\ref{eq:ac_trimer_3} with respect to the intensities $P_j$ ($j=1,2,3$) in each waveguide, as a function of $\mathrm{a}$. The other parameters are kept fixed: $k=0.2$ and $\gamma=1.5$. Solid lines and open circles correspond to the numerically and analytically obtained solutions, respectively. The vertical dashed lines mark the range in which the 
solution $P_1=P_3$ is stable.} 
\label{fig:fp}
\end{figure}

We begin our analysis by seeking stationary solutions. It is easily seen that the origin
$(\psi_1,\psi_2,\psi_3)=(0,0,0)$ is a fixed point of the system. By performing a linear stability analysis around this solution, we can find the characteristic equation analytically:
\begin{eqnarray}
\lambda^3+3\gamma\lambda^2+\lambda(2k^2-2\mathrm{a}^2+3\gamma^2-i4\mathrm{a}k)&& \nonumber\\
-2\mathrm{a}^2\gamma-i4\mathrm{a}\gamma k+\gamma^3+2\gamma k^2=0,&& \nonumber \label{eq:charpoly}
\end{eqnarray}
which gives the following eigenvalues: $\lambda_1=-\gamma-\sqrt{2}\mathrm{a}-i\sqrt{2}k$, $\lambda_2=-\gamma$ and $\lambda_3=-\gamma+\sqrt{2}\mathrm{a}+i\sqrt{2}k$. Therefore, the origin is stable for $a\leq \gamma/\sqrt{2}$ and unstable for $a>\gamma/\sqrt{2}$. For $\gamma=1.5$ and $k=0.2$ this critical value is $a\approx1.06$. This is shown also in Fig.~\ref{fig:fp}, where the intensities in each waveguide $P_j=|\psi_j|^2 (j=1, 2, 3)$ is plotted, as a function of $\mathrm{a}$, while the other parameters $k$ and $\gamma$ are kept fixed. The left dashed horizontal line marks the $\mathrm{a}$ value where the origin loses its stability. 

Beyond this point, the solutions are still stationary but nonzero and have the form
$\psi_j=R_je^{i(\omega z +\phi_j)}$ ($j=1,2,3$), where $R_j$ and $\phi_j$ are independent of the propagation distance. This ansatz leads to the following equations:
\begin{eqnarray}
\omega R_1&=&kR_2\cos{\theta_1}+\mathrm{a}R_2\sin{\theta_1}+{R^3_1}\label{eq:alg1} \\ 
\gamma R_1&=&\mathrm{a}R_2\cos{\theta_1}-kR_2\sin{\theta_1}\label{eq:alg2} \\ 
\omega R_2&=&\mathrm{a}R_1\cos{\theta_1}+kR_3\cos{\theta_2}-\mathrm{a}R_1\sin{\theta_1} \nonumber  \\
&&-\mathrm{a}R_3\sin{\theta_2}+R^3_2 \label{eq:alg3} \\
\gamma R_2&=&kR_1\sin{\theta_1}+kR_3\sin{\theta_2}+\mathrm{a}R_1\cos{\theta_1}\nonumber  \\
&&+\mathrm{a}R_3\cos{\theta_2} \label{eq:alg4} \\
\omega R_3&=&kR_2\cos{\theta_2}+\mathrm{a}R_2\sin{\theta_2}+R^3_3\label{eq:alg5} \\ 
\gamma R_3&=&\mathrm{a}R_2\cos{\theta_2}-kR_2\sin{\theta_2}\label{eq:alg6},
\end{eqnarray}
where $\theta_1=\phi_2-\phi_1$ and $\theta_2=\phi_2-\phi_3$,
and therefore out of six real variables in the system of Eqs~\ref{eq:ac_trimer_1}-\ref{eq:ac_trimer_3} we are left with five representing independent degrees of freedom.

\begin{figure*}[tb]
\begin{center}
\includegraphics[width=\textwidth]{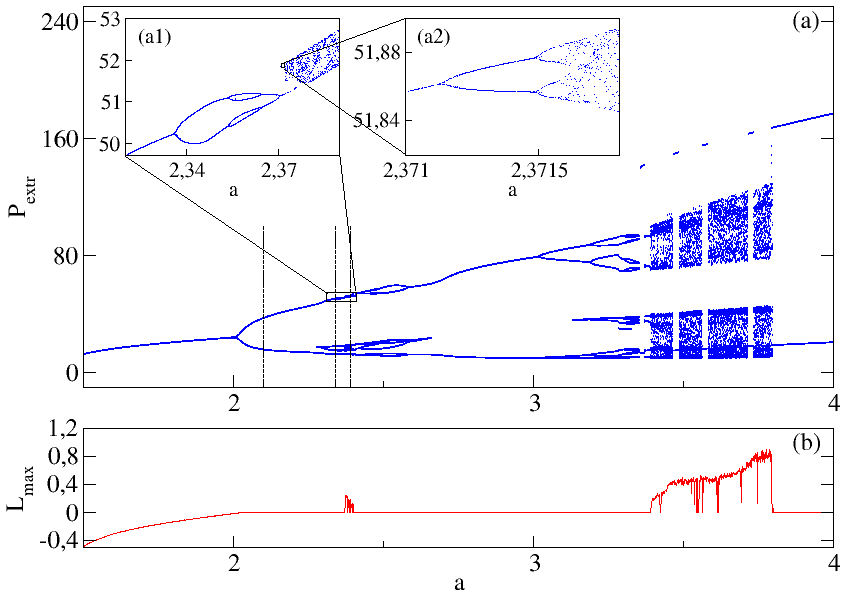}
\caption{(a) Local minima and maxima (extrema) of the total power, $P_{\text{extr}}$, of the AC trimer are plotted against the imaginary part of the complex coupling $\mathrm{a}$ for constant $\gamma=1.5$ and $k=0.2$. The inset (a1) shows a blow-up of (a) and, in turn, inset (a2) shows a blow-up of (a1). (b) The corresponding maximum Lyapunov exponent provides a quantitative measure for the different dynamical regimes.} 
\label{fig:AC_orbit_diagram_LE}
\end{center}
\end{figure*}

From Fig.~\ref{fig:fp}, we see that for the new stationary solution it holds that $R_1=R_3=R$.  In this case, the right hand sides of Eqs.~\ref{eq:alg1} and \ref{eq:alg5} are equal. The same holds for Eqs.~\ref{eq:alg2} and~\ref{eq:alg6} and we therefore obtain:
\begin{eqnarray}
\mathrm{a}\cos{\theta_1}-k\sin{\theta_1}&=&\mathrm{a}\cos{\theta_2}-k\sin{\theta_2}\label{eq:alg7} \\ 
k\cos{\theta_1}+\mathrm{a}\sin{\theta_1}&=&k\cos{\theta_2}+\mathrm{a}\sin{\theta_2}. \label{eq:alg8}
\end{eqnarray}
Multiplying Eq.~\ref{eq:alg7} by $k$ and Eq.~\ref{eq:alg8} by $a$, subtracting them, and vice versa, we get that $\cos{\theta_1}=\cos{\theta_2}=x$ and $\sin{\theta_1}=\sin{\theta_2}=y$. Therefore, given that $R_1=R_3=R$ Eqs.~\ref{eq:alg1}-\ref{eq:alg6} reduce to the
following system of five algebraic equations:
\begin{eqnarray}
\omega R&=&kR_2x+\mathrm{a}R_2y+R^3\label{eq:alg9} \\ 
\gamma R&=&\mathrm{a}R_2x-kR_2y\label{eq:alg10} \\ 
\omega R_2&=&2kRx-2\mathrm{a}Ry+R^3_2\label{eq:alg11} \\ 
\gamma R_2&=&2kRy+2\mathrm{a}Rx \label{eq:alg12} \\ 
1&=&x^2+y^2\label{eq:alg13}.
\end{eqnarray}
Solving Eqs.~\ref{eq:alg10} and~\ref{eq:alg12} with respect to $R_2$, equating them, and
using Eq.~\ref{eq:alg13} we obtain:
\begin{eqnarray}
x&=&\pm \sqrt{\frac{\gamma^2+2k^2}{2(\mathrm{a}^2+k^2)}}\label{eq:sol1} \\ 
y&=&\pm \sqrt{\frac{2\mathrm{a}^2-\gamma^2}{2(\mathrm{a}^2+k^2)}}.\label{eq:sol2}
\end{eqnarray}
%which give us the solutions for $\cos{\theta_1}(\cos{\theta_2})$ and $\sin{\theta_1}(\sin{\theta_2})$ of the initial system of Eqs.~\ref{eq:alg1}-\ref{eq:alg6}.
Moreover, from Eq.~\ref{eq:alg10} we get that:
\begin{eqnarray}
R_2&=&\beta R\label{eq:sol3}\\
\text{with} \hspace{0.2cm} \beta&=&\frac{\gamma}{\mathrm{a}x-ky}\label{eq:sol4}.
\end{eqnarray}
Plugging this into Eq.~\ref{eq:alg9} we can express $\omega$ as a function of $R$, and by substituting it into Eq.~\ref{eq:alg11} we finally obtain:
\begin{equation}
R^2=\frac{2\mathrm{a}y-2kx+\beta(k\beta x+\mathrm{a}\beta y)}{\beta(\beta^2-1)}. \label{eq:sol5}
\end{equation}
To ensure that the right hand side of this expression is positive, the signs of $x$ and $y$ must be chosen correctly. 
From Eqs.~\ref{eq:alg10} and~\ref{eq:alg12} it is easily found that $x$ is positive, therefore the valid sign in Eq.~\ref{eq:sol1} is the ``+" sign. With $x$ being positive, Eq.~\ref{eq:sol5} requires that $y$ is also positive.
To conclude, Eq.~\ref{eq:sol5} (through ~\ref{eq:sol4}) gives us the power in the first and third waveguide $P_1=P_3=R^2$ in dependence of the system parameters $\mathrm{a}$, $\gamma$ and $k$.
This analytic solution together with the corresponding one for $P_2={R_2}^2$ (given by Eq.~\ref{eq:sol3}), are shown with open circles in Fig.~\ref{fig:fp}. The solid lines correspond to the solutions obtained via numerical integration of the initial system (Eqs.~\ref{eq:ac_trimer_1}-\ref{eq:ac_trimer_3}) using a standard fourth-order Runge-Kutta algorithm. The agreement is perfect. This special nonzero stationary solution $P_1=P_3$ is stable up to the value $\mathrm{a}=1.31$, marked by the right vertical dashed line in Fig.~\ref{fig:fp}.

Beyond this critical point, it holds that $R_1 \neq R_3$ and the analytical calculations become more complex. The solutions for $P_1,P_2,P_3$ are derived in the Appendix and are plotted with open circles in Fig.~\ref{fig:fp}: We see that at $\mathrm{a}=1.31$ a supercritical pitchfork bifurcation occurs and two new stationary solutions are born where $P_1$ and $P_3$ are interchangeable: when $P_1$ is on the lower solution branch, $P_3$ is on the higher solution branch, and vice versa. Finally, these two fixed points lose their stability in a Hopf bifurcation at $\mathrm{a}=2.0051$ and two periodic solutions take their place. All this will be discussed in the next section.

%%%%%%%%%%%%%%%%%%%%%%%%%%%%%%%%%%%%%%%%%%%%%%%%%%%%%%%%%%%%%%%%
%%%%%%%%%%%%%%%%%%%%%%%%%%%%%%%%%%%%%%%%%%%%%%%%%%%%%%%%%%%%%%%%

\section{Periodic solutions and Chaos}
\label{sec:chaos}
As we enter the regime of oscillatory solutions, the complexity of the system dynamics increases significantly. 
Figure~\ref{fig:AC_orbit_diagram_LE}(a) shows the orbit diagram in terms of the extrema $P_{\text{extr}}$, i.~e. the maxima and minima, of the total power of light in the trimer $P=|\psi_1|^2+|\psi_3|^2+|\psi_3|^2$ over the imaginary part of the complex coupling coefficient 
$\mathrm{a}$ for a fixed loss rate $\gamma=1.5$ and coupling constant $k=0.2$. The latter is kept constant throughout this section. 

The orbit diagram has been produced via direct numerical integration of our model equations (Eqs.~\ref{eq:ac_trimer_1}\ref{eq:ac_trimer_3}) and scanning of $a$ with a different random initial condition for each realization.

\begin{figure}
%\centering
\includegraphics[width=\linewidth]{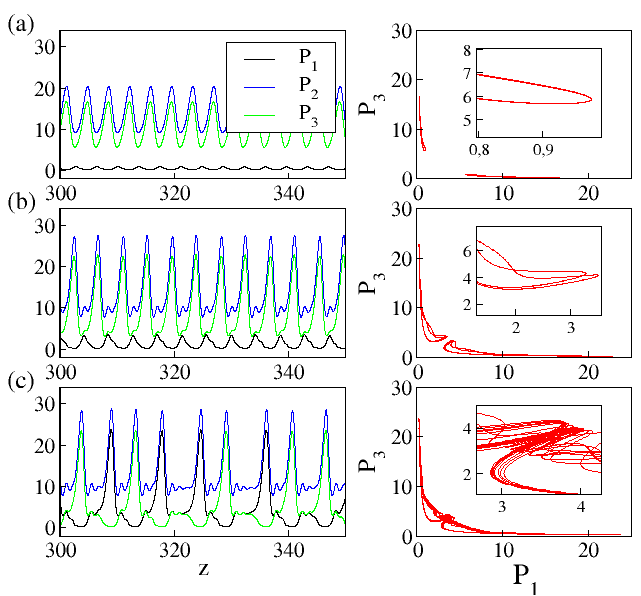}
\caption{Nonlinear dynamics of channel intensities. (a) $\mathrm{a}=2.1$, (b) $\mathrm{a}=2.34$, (c) $\mathrm{a}=2.39$ marked by the dashed vertical lines in Fig.~\ref{fig:AC_orbit_diagram_LE}. The intensity in each waveguide is plotted as we approach chaos in the thin region of Fig.~
\ref{fig:AC_orbit_diagram_LE}. The right panels show the orbits in the $(P_1,P_3)$ plane and the insets in more detail. Other parameters are $\gamma=1.5$ and $k=0.2$.} 
\label{fig:ts_pp}
\end{figure}

As we discussed in the previous section, initially the system's stationary solution is a stable fixed point and the total power is constant. This fixed point loses its stability through a Hopf bifurcation at $\mathrm{a}=2.0051$. Apart from its numerical evidence, the Hopf bifurcation has also been verified by applying a very powerful software tool that executes a root-finding algorithm for continuation of steady state solutions and bifurcation problems~\cite{ENG02}.
At the Hopf bifurcation, a limit cycle is born or, to be more accurate, two coexisting limit cycles are born, which are symmetric with respect to the diagonal in the ($P_1,P_3$) plane and are shown in Fig.~\ref{fig:ts_pp}(a). 

The corresponding total power forms a single oscillatory solution which subsequently undergoes a period-doubling (PD) bifurcation in a thin region, which is blown-up in the inset (a1). The period-2 limit cycle that is born (shown in  Fig.~\ref{fig:ts_pp}(b)) undergoes 
one more PD and subsequently a reverse PD bifurcation, forming thus a stable period-4 ``bubble". The latter is created (destroyed) through a period-doubling (reverse
period-doubling) bifurcation at $\mathrm{a}=2.353$ ($\mathrm{a}=2.364$). The
appearance of bubbles~\cite{BIE84} has been observed in several physical systems such as coupled superconducting oscillators~\cite{HIZ18}.

Eventually, the system enters the chaotic regime through a period doubling cascade shown in the inset (a2). This scenario is repeated on a larger scale in a range of higher $a$ values, and a second wider chaotic regime, including windows of periodic motion within the chaotic attractor, can be observed in the interval $\mathrm{a} \in (3.4,3.8)$. This self-similarity across different scales in the bifurcation pattern is a manifestation of the typical fractal structure of chaos, that we will address later on again.

Complementary to the orbit diagram, Fig.~\ref{fig:AC_orbit_diagram_LE}(b)
shows the maximum Lyapunov exponent $L_\text{max}$ as a quantitative measure for characterizing the previously analyzed dynamical regimes. The maximum Lyapunov exponent has been calculated from the system equations of motion using the method described in~\cite{WOL85}.
As anticipated, there is a perfect agreement between Figs.~\ref{fig:AC_orbit_diagram_LE} (a) and (b), i.~.e. for negative $L_\text{max}<0$ the systems resides in a fixed point (flat $P_\text{extr}$, for $L_\text{max}=0$ the motion is periodic, while for $L_\text{max}>0$ the dynamics is chaotic. Concerning the latter, we observe that the Lyapunov exponents corresponding to the thin chaotic regime (around $\mathrm{a}=2.38$) are much smaller than those of the wide chaotic region ($\mathrm{a} \in (3.4,3.8)$), therefore implying weaker chaos in the former case. 

As mentioned previously, at $\mathrm{a}=2.0051$ just after the Hopf bifurcation we have two limit cycles solutions which are symmetric for the two edge waveguides: Depending on the initial condition the intensity in the first waveguide is low, while the intensity in the third waveguide is high, and vice versa.  Also, the edge waveguides are in anti-phase and the intensity in the middle waveguide is slightly higher compared to that in the edge waveguide. As the system undergoes the cascade of period and reverse period-doubling bifurcations
the two symmetric limit cycles merge into one chaotic attractor (Fig.~\ref{fig:ts_pp} (c)).

%This is very similar to the Lorenz system and the dimer (Barashenkov). Here, additionally we have bubbles! Elaborate on that!

\begin{figure*}
%\centering
\includegraphics[width=\textwidth]{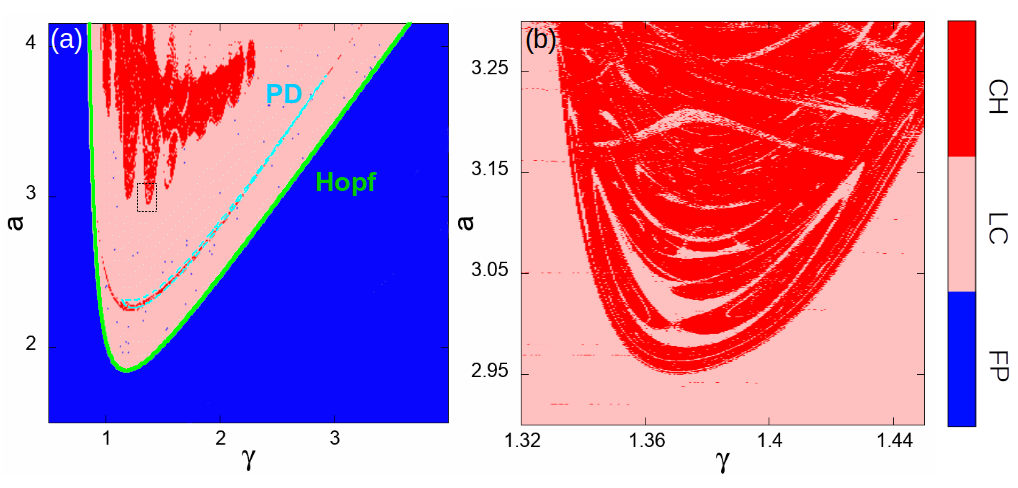}
\caption{(a) The map of dynamical regimes in the $(\gamma,a)$ parameter space. The light green line corresponds to the Hopf bifurcation curve that marks the transition from a fixed point (blue area) to a limit cycle (pink area), while in the red part of the plane the system dynamics is chaotic. The chaotic regime has a an underlying fractal-like structure which is evident in a blow-up of the dotted line rectangle, shown in panel (b).} 
\label{fig:parameter_space}
\end{figure*}
The dynamical scenarios described above are not specific to the selected value of $\gamma$ and $k$ but rather extend to a wider range in the parameter space. For convenience we keep the coupling constant $k$ fixed and explore what happens in the $(\gamma,\mathrm{a})$ plane. In Fig.~\ref{fig:parameter_space}(a) we classify the dynamics of the system based on calculations of the maximum Lyapunov exponent. Blue areas correspond to a negative $L_\text{max}$ and therefore to stationary solutions (fixed points ``FP"), pink corresponds to a zero $L_\text{max}$, i.~.e periodic solutions (limit cycles ``LC"), and red to a positive Lyapunov exponent $L_\text{max}>0$, i.~e. chaotic behavior (``CH"). The border between the blue and pink areas coincides with the Hopf bifurcation line which we have obtained via continuation~\cite{ENG02} and have superimposed on the plot. 
Similarly, we have also included the first period-doubling bifurcation curve that marks the first transition toward chaos occurring in the system.
Figure~\ref{fig:parameter_space}(b) shows a blow-up (marked by the dotted line rectangle) of Fig.~\ref{fig:parameter_space}(a), revealing the fractal-like structure of the chaotic regime that was discussed previously in the context of the orbit diagram Fig.~\ref{fig:AC_orbit_diagram_LE}(a).

%%%%%%%%%%%%%%%%%%%%%%%%%%%%%%%%%%%%%%%%%%%%%%%%%%%%%%%%%%%%%%%%
%%%%%%%%%%%%%%%%%%%%%%%%%%%%%%%%%%%%%%%%%%%%%%%%%%%%%%%%%%%%%%%%

 \section{Conclusions}
We have studied in detail the nonlinear dynamical properties of 
an optical trimer with three lossy waveguides and complex coupling coefficients. The system supports stable stationary solutions, which are obtained analytically as a function of the underlying parameters of the problem, namely the loss, the real and imaginary (gain) parts of the coupling coefficient. In particular, in the oscillatory regime, the system exhibits two coexisting symmetric
limit cycles that undergo a sequence of bubbles and period-doubling bifurcations leading to chaos.
The calculation of the maximum Lyapunov exponent in the gain-loss parameter space reveals a fractal-like structure
of the underlying dynamics. For future studies it would be interesting to study the $\mathcal{PT}$-symmetric trimer
in terms of chaotic behavior. Our preliminary results show that weak chaos is present bur further in-depth studies are required.
%%%%%%%%%%%%%%%%%%%%%%%%%%%%%%%%%%%%%%%%%%%%%%%%%%%%%%%%%%%%%%%%
%%%%%%%%%%%%%%%%%%%%%%%%%%%%%%%%%%%%%%%%%%%%%%%%%%%%%%%%%%%%%%%%
 
 \section*{Acknowledgements}
 This project was funded by the European Research Council (ERC-Consolidator) under grant agreement No. 101045135 (Beyond Anderson).

%%%%%%%%%%%%%%%%%%%%%%%%%%%%%%%%%%%%%%%%%%%%%%%%%%%%%%%%%%%%%%%%
%%%%%%%%%%%%%%%%%%%%%%%%%%%%%%%%%%%%%%%%%%%%%%%%%%%%%%%%%%%%%%%%

\section{Appendix}
Let us consider the general case for Eqs.~\ref{eq:alg1}--\ref{eq:alg6} where $R_1 \neq R_3$.
By setting $x_1=\cos\theta_1$, $y_1=\sin\theta_1$, $x_2=\cos\theta_2$, and $y_2=\sin\theta_2$,
the system equations become:
\begin{eqnarray}
\omega R_1&=&kR_2 x_1+\mathrm{a}R_2 y_1+R^3_1\label{eq:alg1_new} \\ 
\gamma R_1&=&\mathrm{a}R_2 x_1-kR_2 y_1\label{eq:alg2_new} \\ 
\omega R_2&=&kR_1 x_1+kR_3 x_3-\mathrm{a}R_1 y_1-\mathrm{a}R_3 y_2+R^3_2 \label{eq:alg3_new} \\
\gamma R_2&=&kR_1 y_1+kR_3 y_2+\mathrm{a}R_1 x_1+\mathrm{a}R_3 x_2 \label{eq:alg4_new} \\
\omega R_3&=&kR_2 x_2+\mathrm{a}R_2 y_2+R^3_3\label{eq:alg5_new} \\ 
\gamma R_3&=&\mathrm{a}R_2 x_2-kR_2 y_2\label{eq:alg6_new}.
\end{eqnarray}
From Eqs.~\ref{eq:alg2_new} and \ref{eq:alg6_new} we get:
\begin{eqnarray}
R_2&=&\frac{\gamma}{\mathrm{a}x_1-ky_1}R_1 \label{eq:R2_new}\\
R_3&=&\frac{\mathrm{a}x_2-ky_2}{\mathrm{a}x_1-ky_1}R_1. \label{eq:R3_new}
\end{eqnarray}
Substituting Eqs.~\ref{eq:R2_new} and \ref{eq:R3_new} into Eq.~\ref{eq:alg4_new}, we get $x_2$ as a function of $x_1$:
\begin{equation}
x_2=\pm \sqrt{\frac{\gamma^2+2k^2}{\mathrm{a}^2+k^2}-x_1^2}\label{eq:x2}, 
\end{equation}
while:
\begin{eqnarray}
y_1&=&\pm \sqrt{1-x_1^2} \label{eq:y1}\\
y_2&=&\pm \sqrt{1-x_2^2} \label{eq:y2}.
\end{eqnarray}
Moreover, substituting Eqs.~\ref{eq:R2_new} and \ref{eq:R3_new} into Eqs.~\ref{eq:alg3_new} and \ref{eq:alg5_new}, solving 
with respect to $\omega$ and equating both expressions we derive $R_1^2$ as follows:
\begin{equation}
 R_1^2=\gamma\left[1-\left(\frac{\mathrm{a}x_2-ky_2}{\mathrm{a}x_1-ky_1}\right)^2\right]^{-1}\left[ \frac{kx_2+\mathrm{a}y_2}{\mathrm{a}x_2-ky_2}-\frac{kx_1+\mathrm{a}y_1}{\mathrm{a}x_1-ky_1}\right]  \label{eq:R1_new}
\end{equation}
Finally from Eq.~\ref{eq:alg3_new} we find the following expression for $\omega$:
\begin{equation}
 \omega=\gamma \frac{kx_1+\mathrm{a}y_1}{\mathrm{a}x_1-ky_1}+R^2_1. \label{eq:omega}
\end{equation}
All of the derived equations constitute parametric solutions with respect to $x_1$, either directly (Eqs.~\ref{eq:x2}, \ref{eq:y1}, \ref{eq:y2}), or through $x_2$, $y_1$, $y_2$ (Eq.~\ref{eq:R1_new}) and $R_1$ (Eqs.~\ref{eq:R2_new}, \ref{eq:R3_new} and Eq.~\ref{eq:omega}).
Therefore by substituting them into one of the initial Eqs.~\ref{eq:alg1_new}--\ref{eq:alg6_new}, in principle we can find the solution of $x_1$ as a function of the system parameters ($\mathrm{a}$, $k$ and $\gamma$).
However, the calculations involved are extremely complex so we choose to do the following:
We compute the value of $x_1$ as a function of $\mathrm{a}$ via numerical integration of the initial system (Eqs.~\ref{eq:ac_trimer_1}-\ref{eq:ac_trimer_3}), we then calculate $x_2$, $y_1$ and $y_2$ from the algebraic equations Eqs.~\ref{eq:x2}, \ref{eq:y1}, \ref{eq:y2} (we select the ``+" sign in these expressions to ensure that $R^2_1>0$), and by plugging these values into Eq.~\ref{eq:R1_new} we obtain $R_1^2$ as a function of $\mathrm{a}$. Finally, from Eqs.~\ref{eq:R2_new} and \ref{eq:R3_new}, respectively, we also obtain $R_2^2$ and $R_3^2$ as functions of $\mathrm{a}$. These analytical values are shown with open circles in Fig.~\ref{fig:fp}. 

%%%%%%%%%%%%%%%%%%%%%%%%%%%%%%%%%%%%%%%%%%%%%%%%%%%%%%%%%%%%
%%%%%%%%%%%%%%%%%%%%%%%%%%%%%%%%%%%%%%%%%%%%%%%%%%%%%%%%%%%%

\end{document}